\documentstyle[aps,prl,multicol]{revtex}
\begin{document}
\draft
\title{Statistics of soliton-bearing systems with additive noise.}
% \date{\today}
\author{G. Falkovich$^{a,b}$, I. Kolokolov$^{a,c}$, 
V. Lebedev$^{a,d}$, and S. Turitsyn$^e$}
\address{$^{a}$ Physics of Complex
Systems, Weizmann Institute of Science, Rehovot 76100, Israel
\\ $^{b}$ Institute for Theoretical Physics, 
UCSB, Santa Barbara, CA 93106-4030, USA
\\ $^c$ Budker Institute of
Nuclear Physics, Novosibirsk 630090, Russia
\\ $^{d}$ Landau Inst. for Theor. Physics, Moscow, Kosygina 2,
117940, Russia
\\ $^{e}$ Photonics Research Group, 
Aston University, Birmingham B4 7ET, UK}
\maketitle

\begin{abstract}

We present a consistent method to calculate the probability
distribution of soliton parameters in systems with additive noise.
Even though a weak noise is considered, we are interested in 
probabilities of large fluctuations (generally non-Gaussian) which 
are beyond perturbation theory. Our method is a further development
of the instanton formalism (method of optimal fluctuation)
based on a saddle-point approximation in the path integral.
We first solve a fundamental problem 
of soliton statistics governing by noisy Nonlinear
Schr\"odinger Equation (NSE). We then apply our method to optical
soliton transmission systems using signal control elements
(filters, amplitude and phase modulators).

\end{abstract}


\begin{multicols}{2}

Spatially or temporally coherent nonlinear structures (soliton, vortex, 
breather, domain wall, spiral chemical wave, collapsing cavern and many 
others) play an important role in the dynamics and statistics of nonlinear
systems. During the past few decades soliton models have arisen in fields as
diverse as hydrodynamics, plasmas, nonlinear optics, molecular biology,
solid state physics, field theory, and astrophysics. Presumably the most
impressive practical implementation of the fundamental soliton concept has
been achieved in fiber optics, where soliton pulses are used as the
information carriers to transmit digital signal at high bit rates over 
long distances (see e.g. \cite{mmn,NOCN}). 
In long-haul fiber optic communication systems,
the limitations on the bit rate and error-free transmission distance are set
mainly by the spontaneous emission noise added by in-line optical
amplifiers. Existing and
future optical transmission systems can show no measured errors over long
time intervals, that makes a direct modeling of the
bit error rate (that must be less than $10^{-9}$) almost impractical.
An important role is then played by theoretical
methods to evaluate system performance. 
Even though the noise is weak comparatively to the soliton signal,
in general one cannot use perturbation approach to obtain the 
error probability because errors occur when soliton amplitude or
position change substantially due to noise accumulation.
Though dynamical deterministic properties of many nonlinear systems have 
been intensively studied during last decades, much less is known
about statistics of such systems. Typically, consideration of statistical
properties is limited by assumption of the Gaussian statistics and
calculations of the variances (note, however, works \cite{tgnon,cm} where
non-Gaussian corrections due to soliton interaction have been analyzed).
Difficulties in studies of non-Gaussian statistics in
nonlinear systems in particular are caused by lack of appropriate
mathematical methods. In this Letter we present a consistent method to derive
the probability density function in soliton-bearing systems with an additive 
noise. Our approach here is a modification and further development of a 
formalism to calculate the ``probabilities of improbable events''. 
The method is based on finding an optimal fluctuation that provides
for a maximum of probability under given conditions, technically it is 
a saddle-point approximation in the path integral for probability. 
As a specific example, we apply our general scheme to the calculation 
of an error probability in fiber-optic soliton transmission.

We start from the NSE with an additive noise
\begin{eqnarray} &&
-i\partial_t\Psi=\partial_x^2\Psi
+2|\Psi|^2\Psi+\xi \,.
\label{psi1} \end{eqnarray}
Here $\xi$ is white noise with the correlation function
$\langle\xi(t_1,x_1)\xi^*(t_2,x_2)\rangle=
D\,\delta(t_1-t_2)\delta(x_1-x_2)$,
where $D$ is the noise intensity. The equation (\ref{psi1}) 
describes transmission of the signal along the fiber line, then $t$ is 
the propagation distance and $x$ is time.

In this Letter, we focus on the problem of a single soliton 
distortion by the noise. We assume the ideal soliton signal 
$\Psi(0,x)={\cosh^{-1}x}$ at $t=0$ and
examine the probability distribution of different distortions of the
signal at a finite $t=T>0$. Another important problem 
is to find the probability to detect ``one" at a finite distance $T$
provided there was no soliton at $t=0$. Solutions to 
these problems are obtained below by analyzing the noise-induced
fluctuations of $\Psi $ around the ideal soliton form
$\cosh^{-1}x$. We assume that in the soliton units
the distance $T$ is large $T\gg 1$ while the noise power is small 
$D\ll 1$. More precise condition on $D$  will be formulated below. 
To find the probability to lose digital optical 
signal coded by soliton (elementary ``one'') at $t=T$, 
one should define a particular measuring procedure 
that is to specify a receiver. For example, the presence
of the signal at $t=T$ can be established using the value of the
integral $\int_{-l}^l{\rm d}x\,\Psi^\ast(T,x)\Psi(T,x)$. If the window $l$ 
is large enough then the integral is close to $2$ for the soliton 
$\cosh^{-1}x$. Errors are caused by the events with the value of the 
integral essentially smaller than $2$. Such rare events are described by 
the tails of the probability density function (PDF). The focus of our paper 
is to develop a regular method to calculate such (generally non-Gaussian) 
tails of the PDF. In optical applications there are two leading processes 
which can result in these significant (but rare) deviations of the measured 
energy from its mean value. The first process is diminishing of the
soliton power, characterized by the energy integral
$Q=\int_{-\infty}^\infty{\rm d}x\,\Psi^\ast(T,x)\Psi(T,x)/2$,
which is equal to unity for the ideal signal. The second
process is a shift of the soliton position characterized by the
integral $Y=\int_{-\infty}^\infty{\rm d}x\,
x\Psi^\ast(T,x)\Psi(T,x)/(2Q)$ that gives the location of 
the soliton ``center of mass''. For the ideal signal, $Y=0$. It is clear 
that when the soliton almost leaves the detection window $\{-l,l\}$,
the integral $\int_{-l}^l{\rm d}x\,\Psi^\ast\Psi$
one measures can substantially deviate from 2. Therefore 
below we will look for the joint probability distribution function 
${\cal P}(Q,Y)$. 

We parametrize our signal $\Psi$ in the following way:
\begin{eqnarray}
&&\Psi =\eta \exp \left( i\beta x+i\alpha +i\tau \right) 
\left[ \cosh^{-1}(z)+v\right] \,,  
\label{new1} \\
&&z=\eta (x-y)\,,\qquad {\rm d}\tau =\eta ^{2}{\rm d}t\,,  
\label{new2} \end{eqnarray}
where $\alpha ,\beta ,\eta ,y$ are soliton parameters that may be
arbitrary functions of time, and we defined the ``internal time'' $\tau $. 
The field $v$ describes the continuous spectrum of perturbations on 
the background of the soliton. An important observation is that 
at $T\gg 1$ the probabilities of large deviations are determined by
fluctuations of the soliton parameters. This is because the discrete
modes are localized on the soliton and the integral effect of a
continued in time fluctuation can be significant. On the other hand,
dynamics of the field $v$ spread its fluctuations over the whole space. 
We will show below that the influence of the continuous spectrum on 
statistics of the ``soft" variables $\alpha, \beta, \eta$ and $y$ is 
neglibile in the limit $DT^2\ll 1$. The soliton variables themselves
are coupled dynamically in a strong way. We first restrict our
consideration by the set $\alpha, \beta,\eta$ and $y$, then we will
consider the continuous spectrum and establish conditions when it 
can be neglected.

Neglecting the continuous spectrum $v$ we get from (\ref{psi1})
\begin{eqnarray}
&& \partial_t\eta=\eta\zeta\,, \ \  
\partial_t\beta=\zeta_1 \,, \  \ 
\partial_t y-2\beta=\zeta_2 \,.\label{equa} 
\end{eqnarray}
plus an equation for the phase 
$\partial_t\alpha+y\partial_t\beta+\beta^2=\zeta_3$. 
Here the new noises $\zeta(t)\dots\zeta_3(t)$ are some spatial integrals 
of $\xi$. These equations has been derived before, albeit without a careful 
definition of the statistics of the noises \cite{NOCN}. To define the 
statistics, one needs proper regularization of the equations, that 
can be done by considering a limit of a finite correlated case 
$\langle\xi(t)\xi(0)\rangle=g(t)$ with a symmetric function $g$. 
As a result, one may show (the details will be published elsewhere) 
that averaging over $\xi$ is equivalent to the Gaussian average over 
the new noises with  zero cross correlations and the following dispersions:
\begin{eqnarray} &&
\langle\zeta(t)\zeta(0)\rangle\! =\!\delta(t)D/\eta, \ 
\langle\zeta_3(t)\zeta_3(0)\rangle\! =\!\delta(t)(12+{\pi^2}){D}/36{\eta},
\nonumber \\ &&
\langle\zeta_2(t)\zeta_2(0)\rangle =\delta(t){\pi^2D}/{12\eta^3}, \ 
\langle\zeta_1(t)\zeta_1(0)\rangle =\delta(t)D\eta/3\,.
\nonumber \end{eqnarray}
It is interesting that there is a single nonzero average 
$\langle\zeta\rangle ={D}/{2\eta}$ which means a systematic increase of
the soliton amplitude due to the noise: $\langle\eta(T)\rangle=1+DT/2$.

We are going to study phase-independent quantities so $\alpha$ 
variable can be neglected since it neither enters the equations 
for $\eta$, $\beta$, $y$ nor the noise correlators. We may now write 
${\cal P}(Q,Y)$ as an average over the noise of the solutions of the equations
(\ref{equa}) satisfying the boundary conditions $\eta=1$, 
$y=\beta=0$ at $t=0$ and $\eta=Q$, $y=Y$ at $t=T$. Instead of dealing 
with the equations, it is convenient to take the path integral over 
arbitrary functions $\eta$, $\beta$, $y$, taking the equations 
(\ref{equa}) into account by corresponding $\delta$-functions which 
can be rewritten as exponentials introducing auxiliar fields 
$\mu,\mu_1$, $\mu_2$. Performing then averaging over the noise we 
come to the standard Martin-Siggia-Rose formalism:
\begin{eqnarray}
&& {\cal P}(Q,Y)=\int{\cal D}\beta\, {\cal D}\eta {\cal D}y
{\cal D}\mu{\cal D} \mu_1 {\cal D}\mu_2
\exp\left[\int_0^T\!\!{\rm d}t\,{\cal L}(t)\right]\,,  
\nonumber \end{eqnarray}
where the effective Lagrangian is as follows
\begin{eqnarray}  
{{\cal L}}=&&2 i [\mu\partial_t \eta/\eta
-\mu D/2-\mu_1\partial_t\beta
+\mu_2(\partial_ty-2\beta)] 
\nonumber \\ && 
-{D}[12\mu^2+(2\eta\mu_1)^2+(\pi\mu_2/\eta)^2]/(6\eta)
\label{lagr} \end{eqnarray}
Here $\mu(t),\mu_2(t)$ are arbitrary functions on the interval $(0,T)$ 
while $\mu_1(T)=0$ since we do not fix a value of $\beta$ at the final 
moment. From now on we shall be interested in values $|\eta-1|\gg Dt$ 
so we omit the term with the linear drift of $\eta$. 

Since we are interested in the events with small probability, 
we calculate the path integral in the saddle-point approximation:
$\ln{\cal P}\approx \int{{\cal L}}_{\rm saddle}{\rm d}t$. The 
applicability condition of the saddle-point approximation is $DT\ll1$. 
The extrema conditions that determine the saddle-point trajectory 
(also called instanton or optimal fluctuation) can be found from
(\ref{lagr}). Because we are interested in $t\gg1$ one can neglect
the field $\mu_2$ in comparison with $\mu_1$ (as follows from the 
relation $\partial_t\mu_{1}\sim\mu_2$) imposing the corresponding
condition $\partial_ty=0$ at $t=0$. The resulting equations are
\begin{eqnarray}
&&\partial_t\eta=2i D\mu \,, \quad
i\partial_t\mu=D\eta\mu_1^2/3-D\mu^2/\eta \,,
\label{red4} \\
&& i\partial_t^2y=\frac{4}{3}D\eta\mu_1 \,, 
\quad \partial_t^2\mu_{1}=0 \,,
\label{reduce4}
\end{eqnarray}
A solution of (\ref{red4},\ref{reduce4}) is written via Bessel functions:
\begin{eqnarray}
&& \eta=({T-t}) \left[ C_1J_{1/4}(\kappa)
+C_2J_{-1/4}(\kappa)\right]^2 \,,  \label{re3}
\end{eqnarray}
where $\kappa=\lambda(T-t)^2/2$ while the relation of the parameter 
$\lambda$ to $Y$ and $Q$ is found from the boundary conditions. The 
constants $C_1$ and $C_2$ can be expressed via $\lambda$, for instance,
$\sqrt2\,C_2=\Gamma(3/4)\lambda^{1/4}\sqrt{Q}$.
Other fields are easily found now from (\ref{red4}-\ref{re3}).
We are interested here in $\eta\sim1$ (though $1-\eta$ can be much 
larger than its rms value), then  $Y\sim\lambda T^3$.
Therefore, considering a region $Y\ll T$ we get $\lambda T^2\ll1$. 
That procedure corresponds to taking only the first terms 
of the expansion of the Bessel functions in (\ref{re3}). 
Then we get a contribution which is of the second order over
$Y$ (that is Gaussian as a direct result of the
applied perturbation procedure with $\lambda T^2\ll1$):
\begin{eqnarray}
&& \ln{\cal P}(T,Q,Y)\approx -\frac{2}{DT}
\left(\sqrt{Q}\,-1\right)^2 -R(Q) 
\frac{9Y^2}{8DT^3} \,,  
\label{re9} \\
&& R(Q)=10(1+8\sqrt{Q} +Q)(6+3\sqrt{Q}+Q)^{-2}\,.
\label{re10} \end{eqnarray}
This joint energy-timing PDF for the NSE is obtained for the first time.
Remind that it is correct at $Y\ll T$ and $DT^2\ll1$ while $DT^3$ is 
arbitrary. The PDF is Gaussian with respect to timing and non-Gaussian 
with respect to energy. The most important feature of (\ref{re9}) is the 
consistent analytical confirmation of the empirically well-known fact 
that the dispersion of the timing ${4DT^3}/{9}$ is much larger than that 
of the energy ($DT$) so that the error rate of any receiver
with an integration window $l\ll T$ is determined by the timing jitter
due to Gordon-Haus effect \cite{gh,elgin}.
Note that the probability of detecting ``one" formed from noise 
(without any soliton initially present) is given by a
similar instanton solution because an optimal way for a weak noise 
to create a large signal is to grow a soliton. Solving the saddle-point 
equations with the boundary conditions $\eta(0)=0$ and $\eta(T)=\eta_f$ 
we find the probability ${\cal P}(\eta_f)\sim\exp(-{2}\eta_f/DT)$
to observe the amplitude $\eta_f$. Our achievement here is factor $2$.

Next, we examine more sophisticated schemes of optical soliton transmission 
designed to suppress the timing jitter. Their general feature is that to
compensate for the effect of a weak noise
it is enough to modifies the system only slightly so
that an analysis can be done as an extension of the above one.

We consider first the phase modulation which is described by
the equation \cite{smith2}
\begin{equation}
-i\partial _{t}\Psi =\partial _{x}^{2}\Psi 
+2|\Psi |^{2}\Psi +\xi -\epsilon x^{2}\Psi \,,  
\label{ph1} \end{equation}
where the term with $\epsilon$ is regarded to be small.
It produces an additional contribution to the Lagrangian 
$ {\cal L}_\epsilon =4i\epsilon \mu_1{y}$ to be added to 
(\ref{lagr}). Varying the sum ${{\cal L}}+{\cal L}_\epsilon$ 
we get the saddle-point equations 
\begin{eqnarray}
&& \partial_ty=2\beta +iD\frac{\pi^2}{6\eta^3}\mu_2 \,, \qquad
\partial_t\beta=-2\epsilon y -\frac{2i}{3}D\eta\mu_1 \,,  
\nonumber \\
&& \partial_t\mu_1=2\mu_2\,, \qquad 
\partial_t\mu_2=-2\epsilon\mu_1\,,
\label{ph4} \end{eqnarray}
For $\eta$ and $\mu$ we get the same instantonic equations (\ref{red4}).
Below, we will be interested in distances $T\gg1/\sqrt\epsilon$ when the
additional term ${\cal L}_\epsilon$  plays an important role.

One may show that the evolution of $\eta$ and $\mu$ is weakly 
influenced by other degrees of freedom if $y\ll\epsilon^{-1/2}$. 
Then we can develop a scheme similar to that in the basic case: We 
examine first the dynamics of $\eta$ and then the dynamics of $y$ on 
its background as a perturbation. Note that the physics is different now:
The variable $y$ oscillates in a parabolic potential while 
the amplitude of the oscillations grows secularly with time. 
Solving the equations we get for $T>\epsilon^{-1}$:
\begin{eqnarray}
&& \ln{\cal P}(T,Q,Y)\approx -\frac{2}{DT}
\left(\sqrt{Q}\,-1\right)^2
-R_{p}(Q)\frac{3\epsilon Y^2}{DT} \,,  
\label{re91} \end{eqnarray}
with $R_{p}(Q)={3}({1+\sqrt{Q}+Q})^{-1}$. We see that for window 
$l>\epsilon^{-1/2}$ the fluctuations of both timing and amplitude
contribute the error rate, that can be calculated using the 
(substantially non-Gaussian) PDF (\ref{re91}).

Let us discuss now the role of the continuous spectrum. First, it 
is coupled to the discrete degrees of freedom already in the linear 
approximation because of noise. Indeed, if we denote $m$ the field 
conjugated to $v$ (exactly in a way $\mu$ fields are
conjugated to the discrete variables) then the terms proportional to 
$Dm\mu$ appear in the action. A straightforward linear analysis  
shows that $m\ll\mu$ and those mixed terms can be neglected provided 
$T\gg1$. The physical mechanism behind that is the frequency gap 
in the continuous spectrum which makes the mixing nonresonant.
Second, the continuous spectrum influences the soliton parameters 
via nonlinear interaction. The most essential interaction is related to
the terms in the Lagrangian containing $\mu v^2$. Note that the 
fluctuations of the continuous spectrum grow with time,
this effect is related to the noise distributed over the whole space
and is therefore insensitive to the presence of the soliton. Integrating 
over the continuous spectrum, we can find fluctuation corrections to the 
reduced Lagrangian associated with the nonlinear interaction. A relevant 
contribution to the Lagrangian is $\sim(DT)^2 \mu^2$. It has to be compared 
with the bare term $D\mu^2$. Thus we come to the conclusion that our
scheme is valid if $DT^2\ll1$. Note that our analysis of the
continuous spectrum is not sensitive to the presence of the (weak) 
parabolic potential (basically, because of the non-dissipative character of
the phase control). Therefore the criteria $T\gg1$ and $DT^2\ll1$ 
are the same for the phase modulation case.

The most elaborated control scheme that we consider in this Letter
is when filters and amplitude modulators are inserted along the propagation 
line. This scheme is dissipative and it allows one to saturate completely 
the growth of the dispersions of both amplitude $\eta$ and timing 
$y$ with an obvious potential for an unlimited propagation or 
information storage \cite{mmn,NOCN,MMCL}. We analyze below finite 
fluctuations and discover a new (collapse) mechanism of the signal loss 
which restricts the propagation distance or storage time.
The propagation  equation in this case reads (see e.g. \cite{NOCN}): 
\begin{eqnarray} &&
-i\partial _{t}\Psi =\partial _{x}^{2}\Psi 
+2|\Psi |^{2}\Psi +\xi -i\epsilon_{1}\Psi 
-i\epsilon _{2}\partial _{x}^{2}\Psi 
+i\epsilon _{3}x^{2}\Psi \,,
\nonumber \end{eqnarray}
where all $\epsilon$'s are assumed to be small parameters.
The $\epsilon_1$-term describes an additional amplification necessary 
to compensate the losses due to filtering ($\epsilon_2$-term) and 
amplitude modulation ($\epsilon_3$-term). Without noise, one  has a 
steady soliton with $\beta=y=0$ and an amplitude $\eta_s$ satisfying
$\epsilon_1=\epsilon_2\eta_s^2/3 +\epsilon_3\pi^2/12\eta_s^2$. 
It is linearly stable for $4\epsilon_2\eta_s^2>\pi^2\epsilon_3\eta_s^{-2}$, 
the condition is assumed to be satisfied as well as 
$\epsilon_3>2\epsilon_2^2\eta_s^4/9$ that provides for the stability 
of zero \cite{NOCN}. The expression for the dispersions of the energy 
and timing has been derived before and can be found in \cite{NOCN}. 
Here we describe some properties of the whole joint PDF including 
its time-dependent part responsible for a total loss of the signal.
The additional terms produce an additional effect provided 
$T\gg1/\epsilon$, the inequality will be implied below.

An additional contribution to the reduced action 
\begin{eqnarray}
&& \tilde{{\cal L}}_\epsilon
=4i\epsilon_1\mu  -\frac{4}{3}i\epsilon_2\mu\eta^2 -4i\epsilon_2\mu\beta^2
\label{uep2} \\
&&  +\frac{8i}{3}\epsilon_2\mu_1\eta^2\beta 
-2i\epsilon_3\mu\left(\frac{\pi^2}{6\eta^2}+2y^2\right)
- \frac{2\pi^2}{3}i\epsilon_3\mu_2 y/\eta^2   
\nonumber \end{eqnarray}
gives the following saddle-point equations 
\begin{eqnarray}
&& y_t=2\beta -\frac{\pi^2\epsilon_3y}{3\eta^2} 
+\frac{iD\pi^2\mu_2}{6\eta^3} \,, \ 
\beta_t=-\frac{4}{3}
\epsilon_2\eta^2\beta -\frac{2i}{3}D\eta\mu_1 \,,  
\nonumber \\
&& \frac{\eta_t}{\eta}=2\left(\epsilon_1 
-\frac{\epsilon_2\eta^2}{3}-\epsilon_2\beta^2\right) 
-\epsilon_3\left(\frac{\pi^2}{6\eta^2}+2y^2\right)
+\frac{2iD\mu}{\eta}\,. \nonumber
\end{eqnarray}
This system is too complicated to solve it analytically yet the 
most important feature can be understood: If $y(T)=Y$ is sufficiently 
large then the amplitude $\eta$ collapses to zero in a finite time 
according to $\partial_t\eta^2=-\epsilon_3{\pi^2}/{3}$.
For ${\cal P}(Q,Y)$, that means that there is a critical value 
$Y_{{\rm cr}}\sim1$ so that ${\cal P}(Q,Y)$ falls into $\delta(Q)$ if 
$|Y|>Y_{{\rm cr}}$. Of course, $Y_{{\rm cr}}$ 
is a complicated function of $Q$, $\epsilon_1$, $\epsilon_2$, 
$\epsilon_3 $, that can be found only numerically. Below we assume 
all epsilons to be of the same order: 
$\epsilon_1\sim\epsilon_2\sim\epsilon_3\sim\epsilon$.

Let us examine the region of parameters $|Y|<Y_{{\rm cr}}$, $Q\sim1$
that is $y\sim1$ and $1-\eta\sim1$. Then the life time of the corresponding
instanton can be estimated as $\epsilon^{-1}$. Next, we come to estimates 
\begin{eqnarray}
&& \beta\sim\epsilon\,, \quad D\mu_1\sim \epsilon^2 \,, 
\quad \mu\sim\mu_2\sim\epsilon\mu_1 \,.  
\label{uep5} \end{eqnarray}
So, we can conclude that at $|Y|<Y_{{\rm cr}}$ 
the stationary part of the PDF is as follows
\begin{eqnarray}
&& \ln{\cal P}(Q,Y)=-\frac{\epsilon_1^3}{D} F(\epsilon_2/\epsilon_1,
\epsilon_3/\epsilon_1,Q,Y) \,,  \label{pdfmod}
\end{eqnarray}
where $F$ is a dimensionless function of order unity that can be 
found only numerically by finding the extremum of 
$\int_0^T\tilde{{\cal L}}dt$, which is, evidently,  much simpler than 
a massive direct simulations of the noisy NSE.
At $T\gg1/\epsilon$ the PDF decays fast in the region $|Y|>Y_{{\rm cr}}$, 
near the boundary one can estimate
$-\ln{\cal P}(Q,Y)\sim {\epsilon}(Y-Y_{{\rm cr}})^2/D$. 
Thus the region $Y>Y_{{\rm cr}}$ practically doesn't contribute to 
the probability of the signal lost.

The possibility of the collapse leads to the following interesting 
and practically important phenomenon. There is a finite probability per 
unit time for the amplitude to escape the stability region without 
returning. This probability can be found as
a result of the competition of the returning terms in the equation 
for $\eta$ and the noise $\zeta_1$ which indirectly influences $\eta$
through pumping $\beta$ and $y$. The result is the linearly growing 
probability of the total loss of the signal
\begin{eqnarray}
&& {\cal P}_{lost}=T\exp(-F_{{\rm col}}) \,, \quad 
F_{{\rm col}}\sim {\epsilon^3}/{D} \,,  
\label{uep10} \end{eqnarray}
Thus we see that there is a limit time for keeping the information. 

The analysis of the continuous spectrum in this case is slightly different 
due to dissipative character of the additional terms. Therefore a 
saturation of the amplitude is observed which of course is 
$\epsilon$-dependent and tends to infinity when $\epsilon\to0$. 
Estimation of the continuous spectrum fluctuations give a condition
$D\ll\epsilon^{1/2}$ for the above scheme to be valid. Note that the 
conditions of practical applicability is more restrictive:
$\langle Y^2\rangle\simeq D/\epsilon_1\epsilon_2\epsilon_3\ll1$ 
(otherwise the signal will be lost already at $T\simeq 1$).
Note that $D\ll \epsilon^3$ is also the applicability condition 
of the saddle-point approximation as is seen from (\ref{pdfmod}). 

In conclusion, we have developed a consistent mathematical method to 
derive the probability density functions 
in soliton-bearing systems with additive noise.
The method is general and powerful enough to
enable finding probabilities of large deviations in
practical propagation schemes.

G.F. and V.L. thank I. Gabitov for helpful explanations.

\end{multicols}

\end{document}